\documentclass[10pt]{article}

\usepackage{cite} 
\usepackage{amssymb}
\usepackage{epsfig}
\usepackage{graphicx}
\usepackage{psfrag}

\begin{document}

\title{Dirac-Neutrino Magnetic Moment and the Dynamics 
 of a Supernova Explosion}

\author{A. V. Kuznetsov~\footnote{avkuzn@uniyar.ac.ru}, N. V. Mikheev, and A. A. Okrugin\\
\it{Yaroslavl State University, ul. Sovetskaya 14} \\ 
\it{Yaroslavl, 150000 Russia}
}

\date{}

\maketitle

\begin{abstract}
The double conversion of the neutrino helicity $\nu_L \to \nu_R \to \nu_L$ has been analyzed for supernova conditions, 
where the first stage is due to the interaction of the neutrino magnetic moment with plasma electrons and protons 
in the supernova core, and the second stage, due to the resonance spin flip of the neutrino in the magnetic 
field of the supernova envelope. It is shown that, in the presence of the neutrino magnetic moment in the range 
$10^{-13} \, \mu_{\rm B} < \mu_\nu < 10^{-12} \, \mu_{\rm B}$ and a magnetic field of $\sim 10^{13}$\, G between the neutrinosphere and the shock-stagnation 
region, an additional energy of about $10^{51}$\, erg, which is sufficient for a supernova explosion, can be injected 
into this region during a typical shock-stagnation time.
\end{abstract}


Numerical simulations of a supernova explosion 
encounter two main obstacles~\cite{Imshennik:1988,Bethe:1990,Raffelt:1996,Buras:2005,Janka:2007}. 
First, a mechanism 
stimulating the damping shock, which is likely 
necessary for the explosion, has not yet been well 
developed. Recall that shock damping is mainly due to 
energy losses on the dissociation of nuclei. Second, the 
energy release of the ``theoretical'' supernova explosion 
is much lower than the observed kinetic energy 
$\sim 10^{51}$\, erg of an envelope. This is called the FOE (ten to 
the Fifty One Ergs) problem. It is believed that a self-consistent 
description of the explosion dynamics 
requires an energy of $\sim 10^{51}$\, erg to be transferred via 
some mechanism from the neutrino flux emitted from 
the supernova central region to the envelope. 

Dar~\cite{Dar:1987} proposed a possible way for solving the 
above-mentioned problems. His mechanism is based on 
the assumption that the neutrino has a magnetic 
moment that is not very small. Left-handed electron 
neutrinos $\nu_e$ intensely generated in the collapsing 
supernova core are partially converted into right-
handed neutrinos due to the interaction of the neutrino 
magnetic moment with plasma electrons and protons. 
In turn, the right-handed neutrinos sterile with respect 
to weak interactions freely leave the central part of the 
supernova if the neutrino magnetic moment is not too 
large, $\mu_\nu < 10^{-11} \, \mu_{\rm B}$, where $\mu_{\rm B}$ is the Bohr magneton. 
Some of these neutrinos can be inversely converted to 
left-handed neutrinos due to the interaction of the neutrino 
magnetic moment with the magnetic field in the 
supernova envelope. According to contemporary views, 
the magnetic field in this region can be up to about the 
critical magnetic field 
$B_e = m_e^2/e \simeq 4.41 \times 10^{13}$\,G~\footnote{Hereafter, 
we use the natural system of units in which $c = \hbar = 1$, 
and $e > 0$ is the elementary charge.} 
or even higher~\cite{Bisnovatyi-Kogan:1970,Bisnovatyi-Kogan:1989,
Bisnovatyi-Kogan:2005}. The born again left-handed 
neutrinos can transfer additional energy to the envelope by 
virtue of the beta-type absorption $\nu_e n \to e^- p$. 

In our opinion, the mechanism of the double conversion 
of neutrino helicity should be analyzed more carefully. 
It was shown in our recent work~\cite{Kuznetsov:2007} that the flux 
and luminosity of right-handed neutrinos from the central 
region of the supernova were significantly underestimated 
in previous works. Here, we reconsider the process 
$\nu_L \to \nu_R \to \nu_L$ under supernova conditions and 
analyze the possibilities for stimulating the damping 
shock. 

The neutrino spin flip $\nu_L \to \nu_R$ under physical conditions 
corresponding to the central region of the supernova 
has been studied in a number of works (see, e.g., ~\cite{Barbieri:1988,Ayala:1999,Ayala:2000}; 
a more extended reference list is given in~\cite{Kuznetsov:2007}). 
The process is possible due to the interaction of 
the Dirac-neutrino magnetic moment with a virtual 
plasmon, which can be both generated and absorbed: 
\begin{eqnarray}
\nu_L \to \nu_R + \gamma^*, \quad \nu_L + \gamma^* \to \nu_R \, .
\label{eq:conversion}
\end{eqnarray}

In~\cite{Barbieri:1988}, the neutrino spin flip was described in terms of 
scattering by plasma electrons and protons ($\nu_L e^- \to \nu_R e^-$ 
and $\nu_L p \to \nu_R p$, respectively) in a supernova core 
immediately after the collapse. However, the important 
polarization effects of the plasma on the photon propagator 
were not considered in that work. Instead, the 
photon dispersion was taken into account phenomenologically 
by introducing the so-called thermal mass of 
a photon into the propagator. The above-mentioned 
effects were analyzed more consistently in~\cite{Ayala:1999,Ayala:2000},  
where the effect of a high-density astrophysical plasma 
on the photon propagator was taken into account using 
the thermal field-theory formalism. However, an analysis 
of works~\cite{Ayala:1999,Ayala:2000} showed that they concerned only 
the electron component of the plasma, namely, only the 
channel $\nu_L e^- \to \nu_R e^-$, and only the electron contribution 
to the photon propagator, whereas the proton component 
of the plasma was not analyzed at all. This 
seems to be even stranger because the plasma-electron 
and proton contributions to the neutrino spin flip are of 
the same order according to~\cite{Barbieri:1988}.

A consistent analysis of processes~(\ref{eq:conversion}), 
with neutrino-helicity conversion due to the interaction with both 
plasma electrons and protons via a virtual plasmon and 
with the inclusion of polarization effects of the plasma 
on the photon propagator was given in~\cite{Kuznetsov:2007}. 
In particular, 
according to the numerical analysis, the contribution 
of the proton component of the plasma is not 
merely significant, but even dominant. As a result, 
using the data on supernova $SN1987A$, a new astrophysical 
limit was imposed on the electron-neutrino 
magnetic moment: 

\begin{eqnarray}
\mu_\nu < (0.7 - 1.5) \, \times 10^{-12} \, \mu_{\rm B}\,,
\label{eq:mu_fr_Q}
\end{eqnarray}

This is a factor of two better than previous constraints. 
 
In particular, the function $\Gamma_{\nu_R} (E)$ determining the
energy spectrum of right-handed neutrinos was calculated in~\cite{Kuznetsov:2007}. 
In other words, this function specifies the 
number of right-handed neutrinos emitted per 1 MeV of 
the neutrino energy spectrum per unit time from unit 
volume of the central region of a supernova:

\begin{eqnarray}
\frac{\mathrm{d} n_{\nu_R}}{\mathrm{d} E} = 
\frac{E^2}{2 \, \pi^2} \, \Gamma_{\nu_R} (E) \,.
\label{eq:dn/dE}
\end{eqnarray}

In addition, the function $\Gamma_{\nu_R} (E)$ determines the
spectral density of the energy luminosity of a supernova 
core via right-handed neutrinos: 

\begin{eqnarray}
\frac{\mathrm{d} L_{\nu_R}}{\mathrm{d} E}
 = V\, \frac{\mathrm{d} n_{\nu_R}}{\mathrm{d} E} \, E = 
V \, \frac{E^3}{2 \, \pi^2} \, \Gamma_{\nu_R} (E) \,.
\label{eq:dL/dE}
\end{eqnarray}

Here, $V$ is the volume of the neutrino-emitting region. 

The function $\mathrm{d} L_{\nu_R}/\mathrm{d} E$ calculated in~\cite{Kuznetsov:2007} 
is plotted in Fig. 1 for the neutrino magnetic moment 
$\mu_\nu = 3 \times 10^{-13} \, \mu_{\rm B}$. On the one hand, this value is too small to 
affect the supernova dynamics. On the other hand, it is 
sufficiently large to provide the required luminosity 
level. In accordance with the existing supernova models 
(see, e.g., Fig. 11 in~\cite{Sumiyoshi:2005}), a significant part of the 
supernova core material has a fairly high temperature. 
For example, according to the model developed in~\cite{Shen:1998}, 
typical temperatures are 20–-30 MeV. The model proposed 
in~\cite{Lattimer:1991} predicts even higher temperatures. The 
energy distributions of the right-handed neutrino luminosity 
are plotted in Fig.1 for the temperatures $T =$ 35, 
25, 15, and 5 MeV, the electron and neutrino chemical 
potentials $\tilde \mu_e \simeq$ 300 MeV and $\tilde \mu_{\nu_e} \simeq$ 160 MeV, and the 
neutrino-emitting volume $V \simeq 4 \times 10^{18} \, \mbox{cm}^3$. 

\begin{figure}
\begin{center}
\includegraphics*[width=0.75\textwidth]{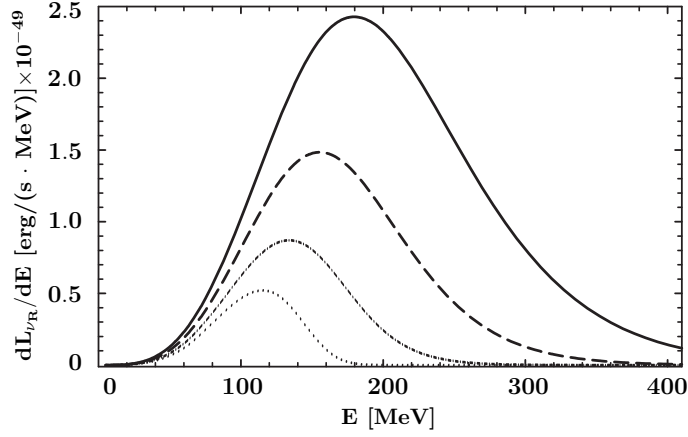}
\caption{Energy distributions of the luminosity of right-handed 
neutrinos for plasma temperatures $T =$ (solid curve) 
35, (dashed curve) 25, (dash-dotted curve) 15, and (dotted 
curve) 5 MeV and the neutrino magnetic moment $\mu_\nu = 3 \times 10^{-13} \, \mu_{\rm B}$.}
\end{center}
\label{fig:emissivity}
\end{figure}

To obtain a total energy of about $10^{51}$ erg extracted 
from the supernova central part by right-handed neutrinos 
in a time of about 0.2 s, the integral luminosity of 
these neutrinos should be about 
\begin{eqnarray}
L_{\nu_R} \simeq 4 \times 10^{51} \, \frac{\mbox{erg}}{\mbox{s}} \,.
\label{eq:L}
\end{eqnarray}
An analysis shows that such luminosity can be generated 
by the considered process of neutrino-helicity 
conversion if the neutrino magnetic moment is not 
larger than refined upper limit~(\ref{eq:mu_fr_Q}) pointed out in~\cite{Kuznetsov:2007}. 
The table illustrates the neutrino magnetic moments for 
which luminosity level~(\ref{eq:L}) is achieved for any of the 
above-mentioned temperatures. 

\begin{table}
\caption{Neutrino magnetic moments providing luminosity~(\ref{eq:L})}
\begin{center}
\begin{tabular}{|c|c|}\hline
$T$ (MeV) & $\mu_\nu / (10^{-12} \, \mu_{\rm B})$  \\ \hline 
35 & 0.29 \\
25 & 0.42 \\
15 & 0.64 \\
5 & 0.97 \\
\hline
\end{tabular}
\end{center}
\end{table}

If the energy of right-handed neutrinos was converted 
into the energy of left-handed neutrinos, e.g., 
due to the well-known mechanism of spin oscillations, 
then an additional energy of about $10^{51}$ erg would be 
injected into the supernova envelope in a typical time of 
about a few tenths of a second. 

It was noted above that the strong dominance of 
neutrino scattering by protons over scattering by electrons 
was not found in earlier studies. Hence, the possible 
number of the right-handed neutrinos generated in 
the collapse of the central region of a supernova was 
significantly underestimated. 

At the same time, it is not evident that neutrino scattering 
by protons dominates over their scattering by 
plasma electrons. In this respect, we believe that a clear 
illustration of this dominance based on an analysis of a 
simplified case is quite expedient. The comparison of 
the typical parameters of the supernova core, where the 
temperature is $T \simeq$ 30 MeV and the electron and neutrino 
chemical potentials are $\tilde \mu_e \simeq$ 300 MeV 
and $\tilde \mu_{\nu_e} \simeq$ 160 MeV, respectively, shows that the temperature is 
the smallest physical parameter. Hence, the limiting 
case of the completely degenerate plasma, $T = 0$, seems 
to yield a reasonable estimate. It is interesting that if the 
temperature is zero, the contributions from neutrino 
scattering by protons and electrons to the neutrino creation 
probability can be evaluated analytically using 
Eqs. (20) and (21) and the corresponding formulas 
from Appendix A in~\cite{Kuznetsov:2007}. The contribution by ultrarelativistic 
electrons is given by the simple formula 
\begin{equation}
\Gamma_{\nu_R}^{(e)}(E) = \frac{\mu_\nu^2\,m_\gamma^2}{2 \, \pi}\,
(\tilde{\mu}_\nu-E) \, \theta (\tilde{\mu}_\nu-E)\,,
\label{eq:zerotlimit_Le}
\end{equation}
where $E$ is the energy of the generated right-handed 
neutrino, $m_\gamma^2 = 2 \, \alpha \, \tilde{\mu}_e^2/\pi$ is the squared mass 
of a transverse plasmon, and $\tilde{\mu}_\nu$ is the neutrino chemical potential. 

The analytical expression describing the proton contribution 
is somewhat more complicated since it 
depends additionally on the proton mass. The plasma 
electroneutrality condition for $T = 0$ takes the form $n_p = n_{e^-}$ 
and ensures that the electron and proton Fermi
momenta are the same: $k^{(e)}_{\rm{F}} = k^{(p)}_{\rm{F}}$. Then, the proton 
chemical potential coinciding with the Fermi energy is
$\tilde{\mu}_p = E^{(p)}_{\rm{F}} =
\sqrt{m_p^2 + \tilde{\mu}_e^2}$ and the proton contribution is 
expressed in terms of the proton Fermi velocity 
$v_{\rm{F}} 
= {k^{(p)}_{\rm{F}}}/{E^{(p)}_{\rm{F}}} 
= {\tilde{\mu}_e}/{\tilde{\mu}_p} 
= {\tilde{\mu}_e}/{\sqrt{m_p^2 + \tilde{\mu}_e^2}}$.
As a result, the proton contribution is given by the expression 
\begin{equation}
\Gamma_{\nu_R}^{(p)}(E) = 
\frac{\mu_\nu^2\,m_\gamma^2\,\tilde{\mu}_\nu}{2\,\pi} 
\, f_p (y) \,, \quad
y = \frac{E}{\tilde{\mu}_\nu} \,.
\label{eq:zl_defns}
\end{equation}
Here, the function $f_p (y)$ has the form 
\begin{equation}
f_p (y) = \frac{1+v_{\rm{F}}/3}{1-v_{\rm{F}}} \; y \, , 
\label{eq:f_y_protons1}
\end{equation}
for $0 \leqslant y \leqslant (1-v_{\rm{F}})/(1+v_{\rm{F}})$ and 
\begin{equation}
f_p (y) = \frac{1-y}{v_{\rm{F}}} \; \theta(1-y) \,  \left[1 - \frac{(1-v_{\rm{F}})^2}{12\,y^2\,v_{\rm{F}}}\,(1-y)\,(1+2\,y)\right]\,.
\label{eq:f_y_protons2}
\end{equation}
for 
$(1-v_{\rm{F}})/(1+v_{\rm{F}}) \leqslant y \leqslant 1$.
It is interesting that the 
integral contribution from protons is independent of the 
parameter $v_{\rm{F}}$: %
$\int_0^1 f_p (y) \, \mathrm{d} y = 1/2$.

Note that the formal passage to the limit $m_p \to 0$
i.e., $v_{\rm{F}} \to 1$ in Eqs.~(\ref{eq:zl_defns})--(\ref{eq:f_y_protons2}) 
yields $f_p (y) \to f_e (y) = (1-y)\; \theta(1-y)$, where the function 
$f_e (y)$ can be introduced in 
Eq.~(\ref{eq:zerotlimit_Le}) in complete analogy with Eq.~(\ref{eq:zl_defns}). Thus, as 
expected, Eq.~(\ref{eq:zerotlimit_Le}) 
for the electron contribution is reproduced. 

Figure 2 shows the plots of the function $f_p (y)$ for 
$v_{\rm{F}}$ = 1, 0.394, and 0. The value $v_{\rm{F}}$ = 0.394 corresponds 
to the effective proton mass $m_p \simeq 700$ MeV in a plasma 
with a nuclear density $3 \times 10^{14}$\, g/cm$^3$ 
(see~\cite{Raffelt:1996}, p. 152). The value $v_{\rm{F}} = 0$ corresponds 
to the formal limit $m_p \to \infty$ 
for which this function is also significantly 
simplified: $f_p (y) \to f_{\infty} (y) = y\; \theta(1-y)$. 

\begin{figure}
\begin{center}
\includegraphics*[width=0.75\textwidth]{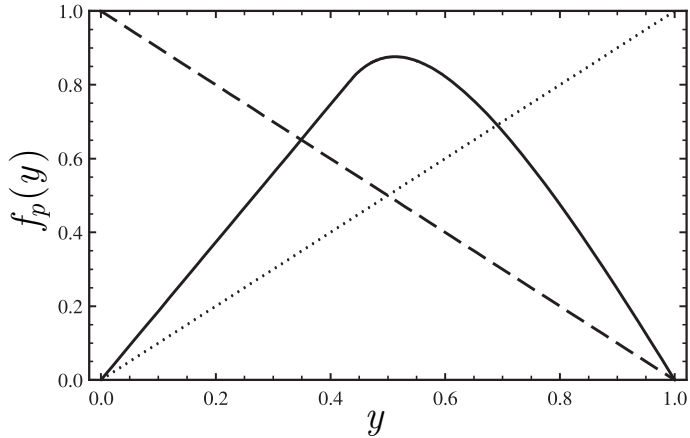}
\caption{Plots of the function $f_p (y)$ for various $v_{\rm{F}}$ values. The 
dependence $f_e (y) = (1-y)$ for the electron contribution is 
reproduced for $v_{\rm{F}} = 1$ (dashed line). The value $v_{\rm{F}} = 0.394$ 
(solid curve) corresponds to the effective proton mass $m_p \simeq 700$
MeV. The case $v_{\rm{F}} = 0$ (dotted line) corresponds to the 
limit of infinitely large proton mass.} 
\end{center}
\label{fig:f(y)}
\end{figure}

According to Eq.~(\ref{eq:dL/dE}), the spectral density of the 
energy luminosity of the supernova core due to right-handed 
neutrinos is given by the formula 
\begin{eqnarray}
\frac{\mathrm{d} L_{\nu_R}}{\mathrm{d} E}
 = V\, \frac{\mu_\nu^2\,m_\gamma^2\,\tilde{\mu}_\nu^4}{4\,\pi^3} 
 \, y^3  
 \left[ f_e (y) + f_p (y) \right] \,.
\label{eq:dL/dE2}
\end{eqnarray}
The difference between the electron and proton contributions 
to the quantity given by Eq.~(\ref{eq:dL/dE2}) is illustrated in 
Fig. 3. It is clearly seen that the proton contribution to 
the luminosity is increased by the factor $y^3$. 

\begin{figure}
\begin{center}
\includegraphics*[width=0.75\textwidth]{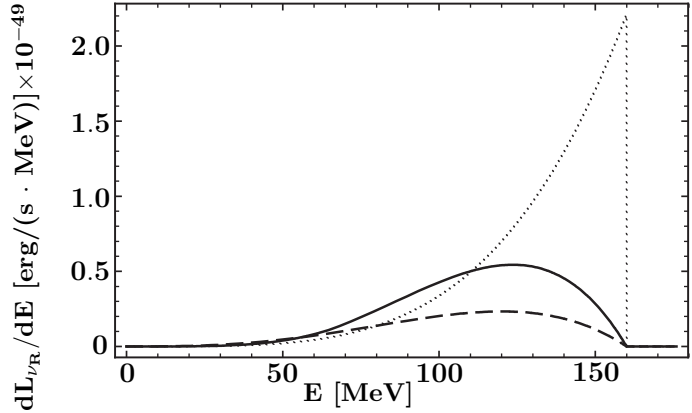}
\caption{Energy distribution of the contributions from 
(dashed curve) electrons and protons with (solid curve) 
$m_p \simeq 700$ MeV and (dotted line) $m_p \to \infty$ to the luminosity 
of right-handed neutrinos for $T = 0$.} 
\end{center}
\label{fig:emissivity_T=0}
\end{figure}

A comparison of Figs. 3 and 1 shows that allowance 
for a nonzero temperature results in a shift of the maximum 
of the energy distribution of the luminosity 
toward higher energies of right-handed neutrinos. This 
additionally enhances the proton contribution. 

The flux of right-handed neutrinos from a collapsing 
supernova core enters the region of the envelope 
between the neutrinosphere of the radius $R_\nu$ and the 
shock-stagnation region of the radius $R_s$. According to 
commonly-accepted notions, the typical values of these 
quantities vary only slightly during the stagnation time 
and can be estimated as $R_\nu \sim$ 20---50 km and 
$R_s \sim$ 100---200 km. If a fairly strong magnetic field of $\sim 10^{13}$ G 
exists in the considered region, then neutrino spin oscillations 
occur and can be resonant under certain conditions. 

The effect of the magnetic field on neutrinos with 
nonzero magnetic moments can be illustrated most 
conveniently using the equation for neutrino-helicity 
evolution in the uniform external magnetic field. The 
helicity-evolution equation taking into account the 
additional energy $C_L$ gained by left-handed electron 
neutrinos in the matter can be written 
as~\cite{Voloshin:1986a,Voloshin:1986b,Okun:1986,Voloshin:1986c,Okun:1988,
Voloshin:1988,Blinnikov:1988} 
\begin{equation}
{\mathrm i}\,\frac{\partial}{\partial t}
\left( 
\begin{array}{c} 
\nu_R \\ \nu_L 
\end{array} 
\right)
=
\left[\hat E_0 +
\left( 
\begin{array}{cc} 
0 & \mu_\nu B_{\perp} \\ \mu_\nu B_{\perp} & C_L
\end{array} 
\right)
\right]
\left( 
\begin{array}{c} 
\nu_R \\ \nu_L 
\end{array} 
\right) \,,
\label{eq:evolution} 
\end{equation}
where 
\begin{equation}
C_L = \frac{3 \, G_{\mathrm F}}{\sqrt{2}} \, \frac{\rho}{m_N} 
\left( Y_e + \frac{4}{3} \, Y_{\nu_e} - \frac{1}{3} \right) \,.
\label{eq:C_L}
\end{equation}
Here, the ratio $\rho/m_N = n_B$ is the nucleon number density 
and $Y_e = n_e/n_B = n_p/n_B, \, Y_{\nu_e} = n_{\nu_e}/n_B$, $n_{e,p,\nu_e}$ 
are the number densities of electrons, protons, and neutrinos, 
respectively, $B_{\perp}$ is the transverse magnetic-field component 
with respect to the direction of neutrino motion, 
and the term $\hat E_0$ proportional to the identity matrix is 
insignificant for our analysis. 

Expression~(\ref{eq:C_L}) for the additional energy of left-handed 
neutrinos is worthy of special analysis. It is 
important that this quantity can vanish in the considered 
region of the supernova envelope. In turn, this is a 
criterion for the resonance transition $\nu_R \to \nu_L$. Since 
the neutrino number density in the supernova envelope 
is fairly low, the quantity $Y_{\nu_e}$ in Eq.~(\ref{eq:C_L}) is negligible.
This yields the resonance condition in the form $Y_e = 1/3$. 
Note that $Y_e$ in the supernova envelope is $\sim$ 0.4–-0.5, 
which is typical of collapsing material. Nevertheless, 
the shock causing the dissociation of heavy nuclei 
makes the material more transparent to neutrinos. As a 
result, the so-called ``short-term'' neutrino burst is generated 
and the material in this region is significantly 
deleptonized. According to conventional notions, a 
characteristic dip down to $\sim$ 0.1 is observed in the radial 
distribution of the quantity $Y_e$ (see, e.g.,~\cite{Bethe:1990,Buras:2005}). The 
qualitative behavior of $Y_e (r)$ is shown in Fig. 4. Thus, a 
point at which $Y_e = 1/3$ certainly exists. It is remarkable 
that only one such point with $\mathrm{d} Y_e / \mathrm{d} r > 0$ exists 
(see~\cite{Bethe:1990,Buras:2005}). 

\begin{figure}
\begin{center}
\includegraphics*[width=0.75\textwidth]{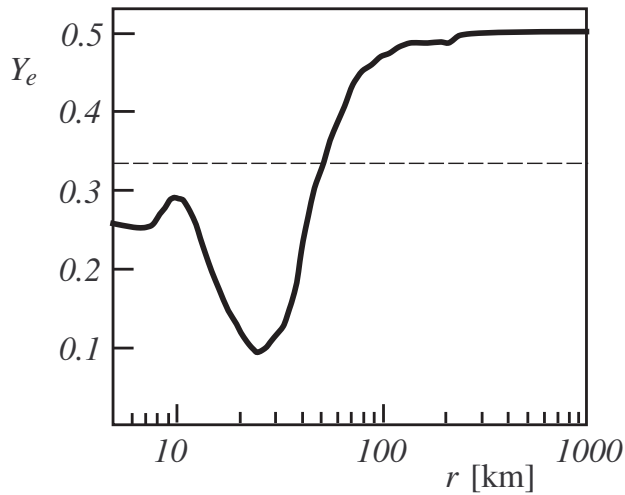}
\caption{Qualitative behavior of the function $Y_e (r)$ at 
0.1–-0.2 s after the shock formation with a typical dip caused by a 
short-term neutrino burst (see, e.g.,~\cite{Buras:2005}). The dashed line 
corresponds to $Y_e = 1/3$.}
\end{center}
\label{fig:gap}
\end{figure}

Note that $Y_e = 1/3$ is a necessary, but insufficient 
condition for the resonant conversion of right-handed 
neutrinos into left-handed ones, $\nu_R \to \nu_L$. 
The so-called adiabatic condition should also be satisfied. The 
meaning of this condition is that the diagonal element 
$C_L$ in Eq.~(\ref{eq:evolution}) should be at least no larger than the 
off-diagonal element $\mu_\nu B_{\perp}$ if the distance from the resonance 
point is about the oscillation length. This leads to 
the criterion~\cite{Voloshin:1988} 

\begin{equation}
\mu_\nu B_{\perp} \gtrsim \left( \frac{\mathrm{d} C_L}{\mathrm{d} r} \right)^{1/2} 
\simeq \left( \frac{3 \, G_{\mathrm F}}{\sqrt{2}} \, \frac{\rho}{m_N} \, 
\frac{\mathrm{d} Y_e}{\mathrm{d} r}\right)^{1/2} .
\label{eq:res_cond}
\end{equation}
The typical parameters of the considered region are 
as follows (see~\cite{Bethe:1990,Buras:2005}): 

\begin{equation}
\frac{\mathrm{d} Y_e}{\mathrm{d} r} \sim 10^{-8} \, \mbox{cm}^{-1} \,, 
\quad
\rho \sim 10^{10} \, \mbox{g} \cdot \mbox{cm}^{-3} \,.
\label{eq:param}
\end{equation}

The magnetic field ensuring the resonance condition 
is 

\begin{equation}
B_{\perp} \gtrsim 2.6 \times 10^{13} \, \mbox{G} 
\left( 
\frac{10^{-13} \mu_{\rm B}}{\mu_\nu} \right) 
\left( \frac{\rho}{10^{10} \, \mbox{g} \cdot \mbox{cm}^{-3}}
\right)^{1/2} 
\left( \frac{\mathrm{d} Y_e}{\mathrm{d} r} \times 10^8 \, \mbox{cm}
\right)^{1/2} .
\label{eq:res_cond_B}
\end{equation}

The mean free path with respect to beta-processes 
for a neutrino with an estimated energy of $E_\nu \sim$ 100---200 MeV is 
\begin{equation}
\lambda \simeq 800 \, \mbox{m} \, \frac{1}{1-Y_e} 
\left( \frac{150 \, \mbox{MeV}}{E_\nu} \right)^2 \, ,
\label{eq:lambda}
\end{equation}
i.e., left-handed neutrinos are almost completely 
absorbed in the considered region. 

Thus, our analysis shows that the Dar mechanism of 
the double conversion of neutrino helicity $\nu_L \to \nu_R \to \nu_L$ 
exists under the following not very severe 
conditions: the Dirac-neutrino magnetic moment should be in the range 
$10^{-13} \, \mu_{\rm B} < \mu_\nu < 10^{-12} \, \mu_{\rm B}$ and a 
magnetic field of $\sim 10^{13}$ G should exist in the region 
$R_\nu < R < R_s$. In this case, an additional energy of about 
\begin{eqnarray}
\Delta E \simeq L_{\nu_R} \, \Delta t \sim 10^{51} \, \mbox{erg} \,,
\label{eq:DeltaE}
\end{eqnarray}
is injected into this region during the shock-stagnation 
time $\Delta t \sim$ 0.2–--0.4 s. This energy is sufficient to solve the 
problem. 


We are grateful to M.I. Vysotsky for helpful discussion. 
This work was supported by the Council of the 
President of the Russian Federation for Support of 
Young Scientists and Leading Scientific Schools 
(project no. NSh-497.2008.2), the Ministry of Education 
and Science of the Russian Federation (Program 
``Development of the Scientific Potential of the Higher 
Education'' (project no. 2.1.1/510), and the Russian Foundation 
for Basic Research (project no. 07-02-00285a).

\end{document}